\documentclass[12pt,a4paper]{article}

\usepackage{amsmath,amssymb,amsthm,graphicx,color,bm,soul}
\usepackage[dvipdfmx]{hyperref} 
\usepackage[nospace]{cite}
\usepackage{leftidx}

\usepackage[normalem]{ulem}
\begin{document}
\title{Coupling quantum-like cognition with the neuronal networks within generalized probability theory} 
\author{Andrei Khrennikov$^{1,*}$,   Masanao Ozawa$^{2,3,4}$,  Felix Benninger$^{5,6}$, \and Oded Shor$^{5,6}$ }
\date{}                     
\maketitle
\noindent
$^{1}$Center for Mathematical Modeling 
in Physics and Cognitive Sciences\\
Linnaeus University, V\"axj\"o, SE-351 95, Sweden\\
$^{2}$Center for Mathematical Science and Artificial Intelligence\\  
Academy of Emerging Sciences, Chubu University, Kasugai, 487-8501, Japan\\
$^{3}$Graduate School of Informatics, Nagoya University, Nagoya, 464-8601, Japan\\
$^{4}$RIKEN Innovation Design Office, Hirosawa, Wako, 351-0198, Japan\\
$^{5}$Felsenstein Medical Research Center, Petach Tikva, Israel\\
$^{6}$Sackler Faculty of Medicine, Tel Aviv University, Tel Aviv, Israel\\
*Corresponding author email: Andrei.Khrennikov@lnu.se

\abstract{The past few years have seen a surge in the application of quantum theory methodologies and quantum-like modeling in fields such as cognition, psychology, and decision-making. Despite the success of this approach in explaining various psychological phenomena — such as order, conjunction, disjunction, and response replicability effects — there remains a potential dissatisfaction due to its lack of clear connection to neurophysiological processes in the brain. Currently, it remains a phenomenological approach.
In this paper, we develop a quantum-like representation of networks of communicating neurons. This representation is not based on standard quantum theory but on {\it generalized probability theory} (GPT), with a focus on the operational measurement framework. Specifically, we use a version of GPT that relies on ordered linear state spaces rather than the traditional complex Hilbert spaces. A network of communicating neurons is modeled as a weighted directed graph, which is encoded by its weight matrix. The state space of these weight matrices is embedded within the GPT framework, incorporating effect-observables and state updates within the theory of measurement instruments — a critical aspect of this model. This GPT-based approach successfully reproduces key quantum-like effects, such as order, non-repeatability, and disjunction effects (commonly associated with decision interference). Moreover, this framework supports quantum-like modeling in medical diagnostics for neurological conditions such as depression and epilepsy.
While this paper focuses primarily on cognition and neuronal networks, the proposed formalism and methodology can be directly applied to a wide range of biological and social networks.}

{\bf keywords:} quantum-like cognition, networks of communicating neurons, directed weighted graphs, generalized probability theory, order effect, interference effect

\section{Introduction}

The quantum information revolution (commonly referred to as the second quantum revolution) has had a profound impact, not only technologically — ushering in advancements in quantum computing, cryptography, and simulation — but also in foundational understanding. One key revelation is that the methodology and formalism of quantum theory can be extended beyond physics to model phenomena in diverse fields, such as cognition, psychology, decision-making, biology (including genetics and epigenetics), evolutionary theory, economics, finance, social and political sciences, and game theory. This interdisciplinary area of research is broadly referred to as quantum-like modeling (e.g., \cite{Zeit}).\footnote{For examples, see monographs \cite{UB_KHR,Busemeyer,Haven,QL3,Bagarello,QLH,Open_KHR, Bagarello1}, the Palgrave Handbook \cite{handbook}, and recent reviews \cite{Pothos,Khrennikov}.} It should be clearly distinguished from the use of genuine quantum physics in cognition, as explored by researchers such as Hameroff \cite{H}, Penrose \cite{P}, and Vitiello \cite{V2}, as well as in biology, particularly in the framework of quantum biophysics \cite{QBIOP, Igamberdiev1}. Quantum-like models, on the other hand, describe the behavior of macroscopic cognitive systems (e.g., human behavior), where the functioning of these systems can be modeled as information processing that adheres to the laws of quantum information theory and probability, deviating from classical laws.

In the past 10–15 years, quantum-like studies have expanded into new domains of science, such as genetics and biological evolution theory \cite{Asano,QIB,BioBas}. As a result, the second quantum revolution has been complemented by the quantum-like revolution \cite{Zeit}. A key challenge for the quantum-like approach lies in the fact that the neurophysiological mechanisms responsible for generating macroscopic quantum-like behavior remain unknown. The experimental statistical data collected in cognitive psychology exhibits phenomena like the order, conjunction and disjunction, response replicability effects, and contextuality. Such data can be successfully described using quantum formalism. \footnote{See, for example, the works cited in \cite{Gabora,B0,Haven2009,Wang,Wang1,PLOS,PO0,Bruza1,OK20,OK21,OK23,Conte,Aerts0,Aerts,Bruza,DZ1,DZ6,CERV,Behti,DBH,Huang,Pothos23,Tsuchiya}.} However, there is still no proper mathematical model that links the neurophysiological processes in the brain with a quantum-like description. This issue was highlighted in \cite{BUSN}.
Currently, this is not merely a theoretical concern. Quantum-like methodology has recently been applied in the medical diagnosis of neurological diseases \cite{Shor1,Shor2}, and the first ``quantum-like patent'' has been granted \cite{Patent}. While this method provides highly accurate predictions, the lack of a neurophysiological foundation hinders its widespread use in hospitals, as physicians require a neurophysiological justification for its application. \footnote{One might refer to quantum biophysics or to genuine quantum physical processing in the brain, such as in the microtubules \cite{H,P}. And the quantum-like approach does not contradict the quantum-brain hypothesis, despite the latter’s numerous drawbacks, as pointed out by many critics, including Tegmark \cite{Tegmark}. From our perspective, the main issue with the quantum-brain theory is that it does not consider neurons as the fundamental units of mental information processing. To identify such units, one would need to explore finer scales of space and time, possibly down to the Planck scale, as in Penrose’s theory of consciousness and quantum gravity. However, we aim to focus on neuronal processing as the core basis for cognition and consciousness.}

In this paper, we propose a mathematical model for the coupling between the brain's neurophysiology and the quantum-like framework. This model is based on the representation of brain neuronal network functioning within {\it generalized probability theory} (GPT) (also known as operational measurement/probability theory) \cite{Mackey,Gudder1,Gudder2,Davies-Lewis,DV,O80,O84,O97,O04}; for reviews, see \cite{Holik, GPTP}. These neuronal networks are considered as systems under analysis. Information exchange between neurons in a network 
is mathematically described by a (generally directed) weighted graph $G_S,$ where the neurons are mapped to the vertices of $G_S.$ 
We acknowledge the challenges in experimentally realizing such mappings for the brain's neuronal networks 
(see, e.g., \cite{Functional}), and thus, our modeling remains far from experimental implementation. Each graph $G_S$ 
 corresponds to the information state of the network, which is described by the weight matrix $\omega=(\omega_{ij}).$ The elements of this matrix encode the strength of signals between communicating neurons, such as the frequencies used in the widely adopted frequency representation of the neural code \cite{Gerstner} or correlations between neuronal activations.
The dynamics of the weighted graph mathematically describes the evolution of the network's states. (Note that the dynamics are not studied in this paper.)

This is an opportune moment to highlight prior research on the neural network foundations of quantum-like models. De Barros and Suppes \cite{DBS} (see also \cite{DB}) demonstrated that neural oscillators can generate interference patterns analogous to those observed in unitary evolution. Takahashi and Cheon \cite{TC} proposed a nonlinear neural network model that uses squared amplitudes to compute response probabilities within a quantum cognitive framework. Busemeyer et al. \cite{BUSN} explored how a neural-based system could perform the computations required for applying quantum probability to human behavior. More recently, quantum-like modeling on treelike graphs has gained traction in medical research \cite{Shor1,Shor2}, reflecting the hierarchical nature of information processing in the brain. Notably, one of the earliest works in quantum-like modeling of cognition \cite{pMENTAL} focused on the quantum mechanics of treelike graphs, which are mathematically represented using $p$-adic numbers. This approach was later extended to quantum-like modeling in studies of petroleum reservoirs \cite{PQ}, where the treelike structure of capillary networks was analyzed. Additionally, in \cite{SR}, mental information was represented quantum-like through the distribution of action potentials in neuron networks.
The present work also draws inspiration from discussions with Gregory Scholes on quantum-like states in neural networks represented by specialized classes of graphs \cite{Scholes1,Scholes2}. However, our approach considers arbitrary graphs.		

This paper is intended for those interested in applying quantum methods to cognition, psychology, and decision-making. Researchers in these areas are generally unaware of GPT. Moreover, GPT is divided into several models that may appear to be different, but are mathematically equivalent \cite{O80} (for reviews, see \cite{Holik, GPTP}). In this paper, we employ a specific model based on {\it ordered linear spaces}, or more precisely, base norm spaces. To make the paper accessible to experts in mathematical psychology, we briefly introduce the basics of GPT within the ordered linear space formalism in section \ref{OLS}.
\footnote{It seems that the GPT approach may be more appealing to psychologists than the standard formalism borrowed from quantum physics. GPT is ``cleansed'' of the quantum mysteries and prejudices, offering a more straightforward general mathematical framework that describes probabilities in all kinds of observations—whether physical, cognitive, social, or financial.}

Before formally introducing GPT, we begin by intuitively exploring and, we hope, clearly using the concepts of GPT. In section \ref{mspace}, we present a model that bridges quantum-like cognition and the brain's networks of communicating neurons. We return to this model in section \ref{OIE} to demonstrate the effects of order, interference (disjunction), and non-repeatability.

This paper is conceptual in nature, aiming to introduce the concept of coupling neuronal activity with the quantum cognition project through GPT. Further papers will delve deeper into the technical details. While this paper focuses on cognition and neuronal networks, the formalism and methodology presented here can be easily applied to a variety of biological and social networks, including quantum-like modeling in genetics, medicine,  and social science \cite{Iryama,Shor1,Shor2,ASR}.

\sloppy
\section{Graph representation of the networks of communicating neurons}
\label{model}

We start with recollection \cite{internet} on the basics of  brain's neuronal signaling: 

\begin{footnotesize}
``The human brain is made up of approximately 86 billion neurons that talk to each other using a combination of electrical and chemical (electrochemical) signals. The places where neurons connect and communicate with each other are called synapses. Each neuron has anywhere between a few to hundreds of thousands of synaptic connections, and these connections can be with itself, neighboring neurons, or neurons in other regions of the brain. A synapse is made up of a presynaptic and postsynaptic terminal. The presynaptic terminal is at the end of an axon and is the place where the electrical signal (the action potential) is converted into a chemical signal (neurotransmitter release). The postsynaptic terminal membrane is less than 50 nanometers away and contains specialized receptors. The neurotransmitter rapidly (in microseconds) diffuses across the synaptic cleft and binds to specific receptors.''
\end{footnotesize}

Voltage-dependent channels endow neurons with active signaling capabilities. Specifically, they enable neurons to generate nerve impulses, also known as {\it action potentials.} Despite variations in duration, amplitude, and shape, action potentials are typically treated as identical events. The duration of an action potential (approximately 1 ms) is often ignored, and an action potential sequence, or spike train, is represented as a series of all-or-none point events. 

It is important to note that each neuron, on average, has about 7,000 synaptic connections with other neurons. Axons vary in length; some are only a millimeter long, while those extending from the brain down the spinal cord can be over a meter in length. An axon has numerous side branches, known as axon collaterals, allowing one neuron to send information to several others. These collaterals, much like the roots of a tree, split into smaller extensions called terminal branches.

Now, consider a network $S$ in the brain consisting of 
$N$ neurons, such as a network performing a specific brain function or even the entire network of neurons in the brain. Neuronal networks are the systems we model in our quantum-like framework. Throughout the paper the symbol  $S$ denotes a neuronal network of communicating neurons.

In the graph representation of $S,$ neurons are mapped to the vertices of the graph, and the connections between neurons are represented as the graph's edges. From an information-theoretic perspective, these edges need not correspond to physical connections, such as axons and synapses. For example, edges (connections) between pairs of neurons may be established based on correlations in their signaling.

We recall that if no two edges share the same pair of endpoints, the graph is said to have no multiple edges. Additionally, if no edge connects a vertex to itself (i.e., no edge has the same vertex as both endpoints), the graph has no loops \cite{graph}. A graph 
$G_S$ representing network $S$ may contain multiple edges and loops.

A directed graph consists of a set of vertices and a set of directed edges, which are ordered pairs of vertices. A directed graph allows loops and permits two edges joining the same vertex but going in opposite directions. However, multiple edges going in the same direction between vertices are not allowed \cite{graph}. The primary model discussed in this article (Model 1, section \ref{mspace}) is based on directed neuronal graphs. The other two models are based on undirected graphs with loops, but without multiple edges.

Graphs describe the geometry of a network. But we model the ``living community'' of neurons communicating with each other. Therefore we consider the weighted graphs. If $G_S$ is the directed graph, then, for each pair of neurons $n_i$ and $n_j$, the weights $w_{ij}$ and $w_{ji}$ are assigned for edges $n_i \to n_j$ and $n_j \to n_i$. If $G_S$ is the undirected graph (without multiple edges), then the weight $w_{ij}= w_{ji}$ is assigned to any edge $(n_i, n_j)=(n_j, n_i)$. 

\subsection{Model 1: directed graph for neurons signaling }

The weight $w_{ij}$ assigned to the graphs' directed edge $n_i \to n_j$ can be interpreted as the probability that during some interval of time $\Delta$ (determining the scale of cognitive information processing in network $S)$, neuron $n_i$ sends a {\it signal}      to neuron $n_j$. (This signal is of the electrochemical nature.) Denote this event $O_{i \to j}$. Thus, the neuronal network  $S$ is represented by the weighted directed graph, each two neurons are connected by two arrow-edges, $n_i \to n_j$ and $n_j \to n_i$ arrows (two neurons cycle), and each neuron $n_i$ is feedback coupled to itself by the loop edge $n_i \to n_i$.  No-signaling is encoded by the zero weight. So, we operate with a complete graph. This is convenient; we don't appeal to the ``hardware'' of the connections between the neurons in the network $S$. By eliminating the edges with $w_{ij}=0$ we would obtain more interesting graph geometry, but this is the topic of further studies.   

\begin{figure}
\begin{center}0,1; 
\includegraphics[width=1\textwidth]{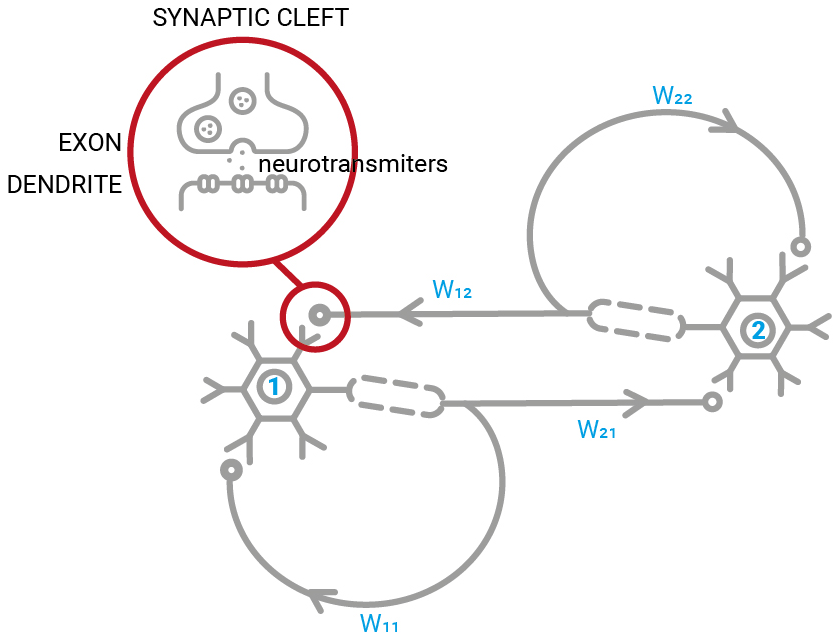}\hspace{8mm} 
\end{center}
\label{Basic}
\caption{An abstract representation of two neurons network.  A directed graph representing neuron 
signaling implies that generally  $W_{12} \neq W_{21}.$ 
Each neuron has two axon collaterals: one ending in a synapse on a dendrite of the other neuron, and the second collateral connecting back to a dendrite of the same neuron. The insert depicts the synaptic cleft.}
\end{figure}

Such neural activity can be represented in the framework of classical probability theory; for example, as follows.  
Let $\xi_{ij}$ denote the classical random variable  
\begin{equation}
\label{xi1}
\xi_{ij}= 
\begin{cases}
    1, & \text{if} \; O_{i \to j} \; \text{holds}, \\
    0,              & \text{otherwise}.
\end{cases}
\end{equation}
Then we set  
\begin{equation}
\label{xi1a}
w_{ij}= E [\xi_{ij}]= P(\xi_{ij}=1).
\end{equation}
Here $0 \leq w_{ij} \leq 1$. We note that generally $w_{ii}>0$ that is a neuron can send signals to itself.  The GPT representation for this neuronal model will be presented in section \ref{mspace}. See Fig. \ref{Basic} for illustration.  
\newpage 

\newpage

\subsection{Model 2: undirected graph for firing neural  code}

Instead of the model with the directed graph, we can consider the simpler model with undirected weighted graph. This model is connected 
with the neural code based on the firing frequency for neurons in the network $S$.  For each neuron  $n_i$ we define the random variable 
\begin{equation}
\label{xi2}
\xi_{i}= 
\begin{cases}
    1, & \text{if} \: n_i \; \text{sends spike},  \\
    0,              & \text{otherwise},
\end{cases}
\end{equation}
that is  in this model we are not interested in  the redistribution of this spike between the neurons of the network $S$, as we do in the directed weighted graph model.  
Then we set  
\begin{equation}
\label{xi2a}
w_{ij}= E [\xi_{i} \xi_{j}]. 
\end{equation}
The weight matrix $W_S=(w_{ij})$ is the covariance matrix of the random vector $(\xi_{1},...,\xi_{N})$.

\subsection{Model 3:  undirected graph for correlation between action potentials}

The third model is also based  on undirected graphs. For each neuron $n_j$ in a network $S$, its action potential is described by the real-valued random variable $\xi_j$.  This is the continuous random variable representing neuron's action potential as a physical potential and not as the discrete spike generation event.  The weights of the graph $G_S$ are given by the pairwise correlations between the action (\ref{xi2a}). The weight matrix $W=(w_{ij})$ is the covariance matrix of the random vector $\xi=(\xi_{1},...,\xi_{N})$. 

We shall construct a GPT representation for Model 1 (section \ref{mspace}). We postpone GPT design for Models 2 and 3 for a future publication.
In the GPT terminology Model 1 has classical state space, but it shows the basic quantum-like effects, as the order, interference (disjunction), and non-repeatability effects.  

\section{GPT for directed neuronal graph (Model 1)}
\label{mspace}

Denote the set of all $N\times N$ real matrices by the symbol $M\equiv M_N$. For  matrix $W=(w_{ij})$,   we define its norm
\begin{equation}
\label{norm2}
||W||=  \sum_{ij} |w_{ij}|. 
\end{equation}
Hence,   $M$ is normed space.   Its dual space $M^\star$, the space of linear functionals, 
is algebraically isomorphic to $M$, so its elements are matrices $F=(f_{ij})$
with the dual paring $\langle F|W\rangle$ such that 
\begin{equation}
\langle F|W\rangle=\sum_{ij}f_{ij}w_{ij}
\end{equation}
for any $W=(w_{ij})\in M$.
The norm on $M^\star$, is given by 
\begin{equation}
\label{f1} 
|||F|||= \max_{||W||=1} |\langle F|W\rangle|=\max_{ij}|f_{ij}|.
\end{equation}
For a matrix $W=(w_{ij})$, we write $W \geq 0$ if, for all $i,j$,  $w_{ij} \geq 0$.  We introduce a partial order structure, $W\geq W^\prime$ if $W - W^\prime \geq 0$. So, $M$ is normed ordered space. A linear functional $F$ is positive, $F \geq 0$,  
if  $\langle F|W\rangle \geq 0$ for all $W \geq 0$, or equivalently
$f_{ij} \geq 0$  for all $i,j$.

The directed weighted graph $G_S$ of neural network $S$ is represented by a matrix
$W_S= (w_{ij}:  w_{ij} \geq 0)$. For example, the matrix elements are given by 
probabilities (\ref{xi1a}). But, other 
neurophysiological entities can also be considered.  To generate dimensionless entities, we normalize these graph-matrices and set
\begin{equation}
\label{norm1a}                                                                                                
\omega\equiv \omega_S=  W_S/||W||.
\end{equation}
Thus, $||\omega_S||=1$.
Denote the set of such matrices by the symbol $\Omega$ and this is the set of the states of the systems under consideration -- the networks of communicating neurons:
\begin{equation}
\label{norm1}                                                                                               
\Omega=\{\omega \in M:  \omega \geq  0, ||\omega||=1 \}.
\end{equation}
We remark that the normalization condition implies that $\omega_{ij} \in [0,1]$.
The set $\Omega$  is convex, closed, and bounded in  $M$  (see Definition 1, section \ref{SO}).    

Now we turn to the issue of pure states, the extreme points of the convex set $\Omega$ (section \ref{SO}). The set of pure states 
coincides with the set of matrices $(E_{ij})$, where $E_{ij}$ is the  matrix with elements  $e_{km;ij}=\delta(k-i) \delta(m-j)$. This is the  basis in linear space $M$ consisting of the pure states. Any state $\omega$ can be represented uniquely as a convex combination (mixture) of pure states
\begin{equation}
\label{mixture}     
\omega= \sum_{ij} \omega_{ij} E_{ij}\, \; \mbox{where}\; \sum_{ij} \omega_{ij} =1, \omega_{ij} \geq 0.
\end{equation}
Thus, $\Omega$ is a simplex. The GPT-models based on such spaces are considered as classical \cite[Theorem II.4.3]{Alfsen}.    
In fact, classical-quantum interplay is more complex and the crucial role is played not by state spaces, but by the calculus of measurement instruments (section \ref{Instruments}).

In GPT observables are defined as the collections of effects, positive linear functionals mapping state space into the segment 
$[0,1]$.  
Any linear functional $F=(F_{ij})$ acts on  a mixture (\ref{mixture}) as
\begin{equation}
\label{functional}     
\langle F|\omega\rangle= \sum_{ij} F_{ij} \omega_{ij},
\end{equation}
so $F$ is an effect (i.e.,  $0\le \langle F|\omega\rangle\le 1$ for all $\omega$) if and only if $0\le  F_{ij}\le 1$ for all $i,j$, or equivalently  $F\ge 0$ and $|||F||| \le 1$.
As in the quantum formalism, for any $W \in M$,  
\begin{equation}
\label{functionala}     
\langle F|W\rangle= {\rm Tr} \; F^\dagger W,
\end{equation} 
where $F^\dagger$ is the adjoint matrix of $F$.
In GPT the important role is played by the unit effect $u$ such that 
$\langle u|W\rangle =||W||$ for $W \geq 0$; see (\ref{L4t}) below).  In the matrix model $u=(u_{ij}=1)$. This is the analog of the trace functional in quantum theory. 

In general this is the basic scheme of GPT, states and effects. To proceed further, one should appeal to theory of measurement instruments. We want to do this properly and this needs more formal mathematical considerations that will be presented in the next section \ref{GPT}.  

\section{Generalized probability theory (operational measurement theory)}
\label{GPT}

GPT was developed in the 1970s as an attempt to understand the foundations of quantum formalism. Within the framework of GPT, the quantum formalism based on complex Hilbert spaces is viewed as a special case of the general operational formalism for all types of measurements. GPT unifies classical and quantum measurement theories and paves the way for a variety of quantum-like formalisms.

It is important to note that the probabilistically non-classical properties of mental phenomena do not necessarily imply that they should be described using quantum formalism (such as the complex Hilbert space), or quantum probability, information, and measurement theory. Some GPT-based models may better align with cognitive information processing and decision-making, as highlighted in \cite{KhrennikovO}. \footnote{In 2015, one of the co-authors (AKH) had extensive discussions in Vienna with Anton Zeilinger about possible applications of quantum theory to cognition. These discussions were prompted by Zeilinger's debates with the Dalai Lama on the relationship between spiritual and quantum mechanical phenomena. Zeilinger strongly advocated for generalized theories that are neither purely classical nor quantum. He also suggested that the ultimate quantum-like theory of cognition might be so mathematically complex and exotic that the complex Hilbert formalism would no longer appear as advanced. However, this is not the case for the model proposed in this article, which is ``less quantum''  (``more classical'') than traditional quantum mechanics.}

In this paper, we employ GPT based on the theory of ordered linear spaces \cite{Davies-Lewis,DV,O80,O84,O97,O04,O23}. Unlike the majority of GPT studies, we focus not on the structures of state and effect spaces, but on the calculus of measurement instruments that provide probabilities (via the generalized Born's rule) and the updates to states generated by the feedback from measurements. This approach, introduced by Davies and Lewis \cite{Davies-Lewis} and known as {\it quantum instruments theory,} has been further developed by Ozawa \cite{O80,O84,O85a,O85b,O86,O97,O04,O23}. A recent article \cite{KhrennikovO} restructured this approach within a contextual framework. The contextual perspective of GPT and quantum instruments theory illuminates their potential applications to cognitive psychology, decision-making, and social science \cite{KhrennikovO}.

GPT research, which focuses on generalizing the concepts of state and observable, was initiated in \cite{Mackey} and significantly developed by Gudder \cite{Gudder1, Gudder2}. Ozawa \cite{O80} established the equivalence between Gudder's and Davies–Lewis’ approaches.
   
\subsection{Ordered linear spaces}
\label{OLS}

For simplicity, we restrict considerations to finite dimensional linear spaces. Let $X$ be a real linear space.
A subset $C$ of $X$ is called a {\it cone,} if for any positive real number $r$, $ rC \subset C$. 
A set $B$ is a {\it base} of the cone $C$ if, for every $x \in C$, there is a unique $\lambda > 0$ such that $\lambda x \in B$ 
or in other words all elements of $C$ can be obtained via scaling of elements of $B$ with positive scalars. 
                                                                         
Let $X$ be a real linear space with a partial order $\leq$ that is consistent with the linear space structure:
for any $x, y,  z \in X$ and real $r \geq 0, $ if $x \leq y$, then  $x+z \leq y+z$ and $rx \leq ry$. Such $X$ is called an {\it ordered linear space.}  Set $X_+=\{x\in X: x \geq 0 \}$. This set is a convex cone containing zero. 
The elements of $X^+$ are called positive. If $x, y \in X$, then $x \leq y$ if and only if $y -x \in X^+$.  

Consider the dual space $X^\star$, the space of linear functionls on $X$. Since we work in the finite dimensional case, spaces $X$ and  
$X^\star$ are algebraically isomorphic.  A linear functional is called positive if 
$f(x) \equiv \langle f|x \rangle \geq 0$ for any $x \geq 0$. The set of all positive functionals $X^\star_+$ is a convex cone 
in $X^\star$.  It is called the dual cone of $X_+$.  The order structure determined by $X^\star_+$ coincides with the point wise order structure on $X^\star$, i.e., $f\leq g$ iff $f(x) \leq g(x)$ for all $x \in X_+$.  
To use analysis, a linear space should be endowed with a topology, e.g., determined by a norm $||\cdot||$  matching the order structure.  Let $X$ be an ordered linear space with norm $||\cdot||$ such that there exists a real number $k > 0$ such that, for all $x, y \in X, 0 \leq x \leq y$ implies that 
\begin{equation}
\label{NOS}
||x||\leq k||y||.
\end{equation} 
Such $X$ is called a {\it normed ordered space.} 
The matrix space $M$ with norm (\ref{norm2}) and its dual space $M^\star$ with norm (\ref{f1}) are normed ordered space with $k=1$. 
 
In any normed space the symbols $B_r(a)$ and $S_r(a)$ denote the ball and sphere 
of radius $r>0$ centered at the point $a$. The dual space is endowed with the norm 
$||f||_\star = \max_{||x|| =1}  |\langle f|x \rangle|$. The corresponding balls and spheres are denoted as $B^\star_r(a)$ and $S^\star_r(a)$. 

\subsection{States and observables}
\label{SO}

{\bf Definition 1.} \cite{GPTP}  {\it State space $\Omega$  is a convex, closed, and bounded subset of a real, finite-dimensional vector space with Euclidean topology.
}\medskip

The above definition is  presented just to simplify the mathematical formalism.  We note that  such set  $\Omega$ is compact.
We adjust this definition to the ordered space approach. Let $X$ be an ordered normed space.
A positive linear functional $f$ on $X$ is called strictly positive if $f(x)>0$ for all 
$ x\in X_+\setminus\{0\}$.
A normed ordered space $X$ is called a {\em base norm space}  if  
$X_+$ is closed and there exists 
a strictly positive linear functional $u$ such that $u(x)=||x||$ for all $x\in X_+$ \cite{Nagel,O80}.
In this case $X_+\cap S_1(0)=\{x\in X_+\mid ||x||=1\}$ is called the {\em  base} of $X$, which is a convex
closed, and bounded base of the cone  $X_+$ of positive elements.
The matrix space $M$ with norm (\ref{norm2}) is a base norm space
but its dual space $M^\star$ with norm (\ref{f1}) is not.

\medskip

{\bf Definition 1a.} {\it 
State space $\Omega$ is 
the base of a finite-dimensional base norm space $X$, i.e.,  $\Omega=X_+\cap S_1(0)$.}

\medskip

Let $\Omega$ be a state space and let $x \in \Omega$. It is said that $x$ is an extreme point of $\Omega$, or
equivalently that $x$ is a pure state, if for every $y, z \in  \Omega$ and $\lambda \in  (0; 1)$ such that 
$x =\lambda  y + (1 - \lambda )z$ we have $x = y =z$. And a state space $\Omega$ is a {\it polytope} if it 
has finitely many pure states.
 
 A positive functional, $E \in X^\star_+$, is called an {\it effect} on the state space $\Omega$ if 
\begin{equation}
\label{L4}
0 \leq \langle E | \omega \rangle \leq 1 \; \mbox{for \;all} \; \omega \in \Omega.
\end{equation}
Effects are the basic observables in the operational theories of measurement. 
They can be described solely in terms of functions on the state space  $\Omega$, as linear space extensions of affine functionals defined on $\Omega$ and  valued in the segment $[0, 1]:$ 
\[\langle f| \lambda \omega_1 + (1-\lambda) \omega_2 \rangle= \lambda \langle f|\omega_1\rangle + (1-\lambda) \langle f| \omega_2\rangle \] 
for $\lambda \in [0,1]$ and $\omega_1, \omega_2 \in \Omega$.
The set of effects is denoted by the symbol ${\mathcal E}$.  This set is convex and compact.  
It follows from Definition 1a
 that ${\mathcal E}$ contains the unit effect $u$, for $x \in X_+$,
\begin{equation}
\label{L4t}
\langle u|x\rangle=1 \; \mbox{if and only if}\;  x \; \mbox{belongs to} \; \Omega,
\end{equation}
that is $\langle u|x\rangle=||x||$ for all $x\in X_+$ and
\begin{equation}
\label{L4ta}
\Omega=\{ x: x \geq 0, \langle u|x\rangle=1 \}\; 
\end{equation}
  In terms of the unit functional, 
states can be characterized as positive vectors constrained by equality  (\ref{L4t}). 

 Let $(E_j)$ be a sequence of effects such that 
\begin{equation}
\label{L5}
\sum_j E_j = u, \; \text{that is} \; \sum_j \langle E_j |\omega \rangle =  \langle u |\omega \rangle =1,  
\end{equation}
and let $(\alpha_j)$ be a sequence of real numbers.   An {\it observable} $A$ is a sequence of couples
$(\alpha_j, E_j)$, where set $ V_A= (\alpha_j)$ encodes the outcomes of observations (``spectrum'' of the observable) and the effects determine the probabilities 
of these outcomes via the generalized {\it Born's rule}:
\begin{equation}
\label{L6}
P(A=\alpha_j|\omega) = \langle E_j| \omega\rangle,  \; \omega \in \Omega. 
\end{equation}  
The average outcome of observable $A$ in the state $\omega$ is as usual via the formula:
\begin{equation}
\label{L5a}
\langle A\rangle_\omega= \sum_j \alpha_j P(A=\alpha_j|\omega)=  \langle \sum_j \alpha_j E_j| \omega\rangle.
\end{equation} 
Thus, formally the observable $A$ can be described as the functional $A= \sum_j \alpha_j E_j \in X^\star;$ then
$\langle A\rangle_\omega= \langle A|\omega\rangle$. But this description is useful only for calculation of averages  

\subsection{Instruments}
\label{Instruments}

Let $J: X \to X$ be a linear operator preserving the positive cone, i.e., 
$J(X_+) \subset X_+$.    Its dual operator $J^\star: X^\star \to X^\star$ also preserves positivity:
let $x \in X_+,  f \in X^\star_+$, then $\langle J^\star f|x\rangle= \langle  f|J x\rangle\geq 0$, since $ J x\geq  0$.   
 
Let $A=(\alpha_k, E_k)$ be an observable with the discrete set of outcomes $ V_A=(\alpha_k)$ and the family of effects $(E_k)$ 
and let, for each outcome $\alpha\in  V_A$, there is defined a linear operator 
$ J(\alpha): X \to X$ preserving the positive cone $X_+$. 
It is assumed that the family of maps $( J(\alpha))$ is constrained as
\begin{equation}
\label{L8}
 J( V_A)\equiv \sum_\alpha  J(\alpha): \Omega \to \Omega.
\end{equation} 
We remark that this constrain implies that
\begin{equation}
\label{L8j}
 \langle u|  J( V_A) x\rangle = \langle u|  x\rangle, \; x \in X_+.
\end{equation} 
For $x \in X_+$, set $\omega_x= x/ \langle u|  x\rangle$. Since $\langle u|  \omega_x \rangle=1$, we have that 
$\omega_x \in \Omega$ and, hence, $\langle u|  J( V_A) \omega_x\rangle= 1$, and (\ref{L8j}) holds. 

Hence, 
\begin{equation}
\label{L8tu}
0 \leq \langle u|  J(\alpha)\omega\rangle \leq 1
\end{equation}
We remark that the last part of equality (\ref{L8j}) implies that 
\begin{equation}
\label{L8jb}
\langle u|  J( V_A) \omega\rangle =1.
\end{equation}

In operational measurement theory, measurements with the outcome $\alpha$  induce the state-update transform of the state space $\Omega:$
\begin{equation}
\label{L7}
\omega \to \omega_{\{A=\alpha\}} = \frac{ J(x)\omega}{\langle u|  J(\alpha)\omega\rangle}.
\end{equation} 
We remark that $\langle u| \omega_{\{A=\alpha\}} \rangle=1$, hence $ \omega_{\{A=\alpha\}}\in \Omega$, see (\ref{L4t}).

For each $ J(\alpha): X \to X$ consider the dual operator $ J^\star(\alpha): X^\star \to X^\star;$ it is positivity preserving.
Hence, $ J^\star(\alpha) u \in  X^\star_+$. Moreover, $\langle  J^\star(\alpha) u | \omega\rangle = 
\langle u |  J(\alpha)  \omega\rangle$. Hence, 
\begin{equation}
\label{L9h}
\sum_\alpha \langle  J^\star(\alpha) u | \omega\rangle = \langle u | \sum_x  J(\alpha)  \omega\rangle=
\langle u |  J( V_A)\omega\rangle=1.
\end{equation}
Thus, for any outcome $x, \;  0 \leq \langle  J^\star(\alpha) u | \omega\rangle \leq 1$. 
The functional 
\begin{equation}
\label{L9}
E(\alpha)=  J^\star(\alpha) u
\end{equation}
 is an effect. Moreover, equality (\ref{L9h}) implies that, for any $\omega \in \Omega$, 
\begin{equation}
\label{L9w}
 \langle \sum_\alpha E(\alpha) | \omega\rangle  =
\sum_\alpha \langle  J^\star(\alpha) u | \omega\rangle=1 .
\end{equation}
Since $u$ is uniquely characterized by the property, $\langle u| \omega\rangle =1$ for any $\omega \in \Omega$,  equality (\ref{L9w})  implies that $\sum_\alpha E(\alpha)=u$, i.e., the system $A=(\alpha, E(\alpha))$ is an observable. 

{\bf Definition 2.}  {\it A family of positivity preserving linear operators $ J(\alpha): X \to X, \; \alpha \in  V_A$, constrained as (\ref{L8}) is called a measurement instrument.} 

Each  instrument generates an observable $A=(\alpha,E(\alpha))$ with the 
effects given by the equality (\ref{L9}); the probabilities of outcomes are determined by the generalized Born's rule:
\begin{equation}
\label{L9d}
P(A=\alpha|\omega)= \langle E(\alpha)|\omega \rangle =  \langle u|  J(\alpha)\omega \rangle,
\end{equation}
measurement's back action on a state $\omega$ is determined by  formula (\ref{L7}). 

{\bf Definition 3.}  {\it Generalized probability theory (GPT), or operational measurement theory,  is a triple $(X, \Omega,  \mathfrak{J})$, 
where $X$ is a  base norm space, $\Omega$ is a state space, and  $ \mathfrak{J}$ is 
the set of measurement instruments generating observables (families of effects) by the rule (\ref{L9}).}

As emphasized in \cite{OK21}, the foundation of any measurement model lies in its instrument calculus. One cannot start with observables, as is often done in quantum information theory, where the introduction of observables is typically sufficient. In reality, an observable is merely a byproduct, or an ``effect,'' of an instrument, and the statistics of a measurement are fully described by the corresponding instrument.

\subsection{Conditional and sequential joint probability distributions}

Within GPT, operational measurement theory, we can define conditional probability \cite{KhrennikovO,O85a,O85b}. Let $( J_A(\alpha))$ and $( J_B(\beta))$ be two instruments generating the observables $A$ and $B$. The probability to obtain the output $A=\alpha$ under the condition  that the output $B=\beta$ was obtained in the preceding measurement of observable $B$ is given by the formula:
\begin{eqnarray}
\label{L9j}
P(A=\alpha|B=\beta, \omega)&=&P(A=\alpha|\omega_{\{B=\beta\}})
= \langle E_A(\alpha)|\omega_{\{B=\beta\}} \rangle\nonumber\\
&=& \langle u|  J_A(\alpha) \omega_{\{B=\beta\}} \rangle
=\frac{\langle u|  J_A(\alpha)  J_B(\beta) \omega \rangle}{\langle u|  J_B(\beta) \omega \rangle}\nonumber\\
&=&
\frac{\langle u|  J_A(\alpha)  J_B(\beta) \omega \rangle}{P(B=\beta| \omega)} .
\end{eqnarray}
By using conditional probability we define the sequential joint probability  
\begin{equation}
\label{L9k}
P_{BA}(\beta,\alpha|\omega)=P(B=\beta|\omega) P(A=\alpha|B=\beta, \omega)= \langle u|  J_A(\alpha)  J_B(\beta) \omega \rangle.
\end{equation}
\sloppy
Sequential measurement of observables $A_1,..., A_n$ corresponding to instruments $ J_{A_j}$ generates the joint probability distribution 
$
P_{A_1...A_n}(\alpha_1,..., \alpha_n)= \langle u|  J_{A_n}(\alpha_n)... J_{A_1}(\alpha_1)  \omega \rangle.
$

\subsection{Order effect}
\label{OE}

The question order effect is well-studied in psychology and sociology, particularly in the context of social opinion polls \cite{Moo02}. In classical probability theory, the order effect is absent. In \cite{Wang, Wang1}, this effect was represented within a quantum-like framework based on projection-type (von Neumann \cite{VN}) state updates. However, in \cite{PLOS}, it was demonstrated that such a projection-based description cannot be combined with another psychological effect, the response replicability effect. To address this issue, the calculus of quantum instruments was employed in \cite{OK20, OK21}. We now examine the order effect within the framework of GPT.

Consider two observables (in psychology - questions, tasks) $A, B$ associated with the instruments  $ J_{A}, J_{B}$. They show {\it  the order effect} in a state $\omega$, if for some outcomes $\alpha_1$ and $\alpha_2$, 
\begin{equation}
\label{L9o}
P_{AB}(\alpha, \beta|\omega) \not= P_{BA}( \beta, \alpha|\omega),
\end{equation} 
that is 
$$
\langle u| ( J_{B}( \beta) J_{A}(\alpha) -  J_{A}(\alpha) J_{B}( \beta)) \omega \rangle \not=0,
$$ 
that is the instrument operators don't commute in this state (cf. \cite{OK21}),
\begin{equation}
\label{L9a}
[ J_{A}(\alpha),  J_{B}( \beta)] \omega \not=0.
\end{equation}

\subsection{Interference effect}
\label{IE}

We note that probability interference is one of the fundamental quantum effects. For example, this effect can be observed in the double-slit experiment, as emphasized by Feynman \cite{F1,F2}, and further discussed in \cite{INT}. In cognitive science and decision-making, it was analyzed in \cite{QL0} as a nonclassical property of cognitive information processing. Its connection with the disjunction effect in cognitive psychology was established in \cite{B0}. It is important to note that in classical probability theory, probabilities do not interfere.

As was explained in \cite{INT}, quantum-like interference of probabilities can be written as the violation of one of the basic laws of classical probability theory, the formula of total probability (FTP). For dichotomous observables $A=\alpha_1, \alpha_2$ and $B=\beta_1, \beta_2$, FTP has  the form:
$
P(A=\alpha)= \sum_\beta P(B=\beta) P(A=\alpha|B=\beta).
$
In classical probability the role of the state is played by the probability measure of the Kolmogorov probability space, to match our formalism denote this Kolmogorovian probability by the symbol $\omega$, then FTP is written as
\begin{equation}
\label{L9m}
P(A=\alpha|\omega)= \sum_\beta P(B=\beta|\omega) P(A=\alpha|B=\beta,\omega).
\end{equation}
In an operational measurement model the left-hand side of FTP equals   
$
P(A=\alpha|\omega)= \langle u|  J_{A}(\alpha)  \omega \rangle
$
and the right-hand side  of FTP equals 
\begin{eqnarray}
\lefteqn{\sum_\beta P(B=\beta|\omega) P(A=\alpha|B=\beta,\omega)}\qquad\nonumber\\ 
&=& \sum_\beta P_{BA}(\beta,\alpha|\omega)
=\sum_\beta \langle u|  J_A(\alpha)  J_B(\beta) \omega \rangle\nonumber\\
&=& \langle u|  J_A(\alpha)  J_B( V_B) \omega \rangle.
\end{eqnarray}
The interference term is defined as the difference between these two expressions
\begin{equation}
\label{L9ma}
\delta_{BA,\omega}(\alpha)= \langle u|  J_{A}(\alpha) (I-   J_B( V_B))\omega \rangle.
\end{equation}
If $\delta_{BA,\omega}(\alpha)=0$ we have classical FTP (for this $\beta)$, if $\delta_{BA,\omega}(\alpha)>0$, this is constructive interference, it increases the probability (this is the basic tool of increasing effectiveness of quantum computing comparing with classical one), if $\delta_{BA,\omega}(\alpha)<0$, then interference is destructive.  

Suppose that 
\begin{equation}
\label{L9mb}
[ J_{A}(\alpha) ,  J_B( V_B)] \omega =0,
\end{equation}
in this case 
$
\langle u|  J_{A}(\alpha)    J_B( V_B)\omega \rangle=  
\langle u|    J_B( V_B)   J_{A}(\alpha) \omega \rangle .
$
Since $ J_{A}(\alpha) \omega \geq 0$, by equality (\ref{L8j}) $\langle u|    J_B( V_B)   J_{A}(\alpha) \omega \rangle
=\langle u|    J_{A}(\alpha) \omega \rangle$. Therefore instrument commutativity condition (\ref{L9mb}) implies that the interference term  is zero, $\delta_{BA,\omega}(\alpha)=0$. 

\subsection{Repeatable measurements}
\sloppy
Consider observable $A=(\alpha, E_\alpha)$ based on instrument $ J$  such that for some $a$, 
$J(a)$ is a positive projection that is $J(a) \geq 0$ and $J(a)^2=J(a)$. Then, for any state $\omega$, 
$P(A=a|A=a, \omega)=P(A=a|\omega_{\{A=a\}}) = \langle u| J(a) \omega_{\{A=a\}}\rangle = 
\langle u| J(a)^2 \omega\rangle/\langle u| J(a) \omega\rangle =1$. We also have $ J( V_A)= \sum_\alpha  J(\alpha): \Omega \to \Omega$.
Thus,  $\sum_\alpha  P(A=\alpha|A=a, \omega) =  \langle u|   J( V_A) \omega_{\{A=a\}}\rangle=1$. Consequently, 
\begin{equation}
\label{h}
P(A=\alpha|A=a, \omega)= \delta_{\alpha,a}
\end{equation}
Hence, for the $a$-outcome the measurement is repeatable.  However, if, for some $a,  J(a)$ is not projection, i.e., 
$J(a)^2\not= J(a)$, then generally $P(A=a|A=a, \omega)= \langle u| J(a)^2 \omega\rangle/\langle u| J(a) \omega\rangle \not= 1$. 

We remark that in classical probability theory all measurements (they are given by random variables) are repeatable; by the Bayes formula for conditional probability, 
$P(A=\alpha|A=a, \omega)= P(A=\alpha, A=a| \omega)/ P(A=a| \omega)= \delta_{\alpha,a}$. We also remark that in quantum probability theory 
the projection type measurements are repeatable. 

\section{Model 1: Order, interference, and repeatability effects} 
\label{OIE}

In matrix space $M_2$ consider observables $A$ and $B$ with the values $\alpha_1, \alpha_2$ and $\beta_1, \beta_2$ with instruments realized as multiplication by matrices: 
$$ 
J_A(\alpha_1) = \begin{vmatrix}
1&0\\
0&0\\
\end{vmatrix},  J_A(\alpha_2) =\begin{vmatrix}
0&0\\
0&1\\
\end{vmatrix} ;  J_B(\beta_1) =\begin{vmatrix}
0&1\\
0&0\\
\end{vmatrix} ,  J_B(\beta_2) = \begin{vmatrix}
0&0\\
1&0\\
\end{vmatrix},
$$
then $J_A(V_A)= \begin{vmatrix}
1&0\\
0&1\\
\end{vmatrix}, \; J_B(V_B) =\begin{vmatrix}
0&1\\
1&0\\
\end{vmatrix}$ and, for $\alpha_1$ and $\beta_1$, we have
$J(\alpha_1) J(\beta_1)= J(\beta_1), J(\beta_1) J(\alpha_1)=0$ that is 
$[J(\alpha_1), J(\beta_1)]= J(\beta_1)\not=0$. 

For an arbitrary state $\omega,\; 
P_{BA}( \beta_1, \alpha_1)=\langle u|  J(\alpha_1) J(\beta_1) \omega\rangle= \langle u|  J(\beta_1)  \omega\rangle=
\omega_{21}+ \omega_{22} \not= P_{AB}(\alpha_1,  \beta_1)= 0$. Thus, the order effect can be found for 
any network  of two neurons $(n_1,n_2)$ in that $n_2$ is active.  

Now we turn to the interference effect (in psychology it is coupled to the disjunction effect); for $\alpha_1$ the interference term$,\delta_{BA,\omega}(\alpha_1)= \langle u|  J_{A}(\alpha_1) (I-   J_B( V_B))\omega \rangle$. We note that 
$J_{A}(\alpha_1) (I-   J_B( V_B))=\begin{vmatrix}
1&-1\\
0&0\\
\end{vmatrix}$. Thus, $\delta_{BA,\omega}(\alpha_1)= (\omega_{11}+ \omega_{12}) - (\omega_{21} +\omega_{22})$. This term equals zero iff the average outputs of neuron $n_1$ and $n_2$ are equal.

Now turn to the problem of repeatability of observations. We note that $J(\beta_1)^2=0$, i.e., it is not projection; then 
$P(B=\beta_1|B=\beta_1, \omega)= 0$. Therefore $B$ measurement is not repeatable. In contrast the $A$-measurement is repeatable.  

\section{Model 1: Compound systems, tensor product representation}
\label{compound}

Consider now two neuronal networks $S_k, k=1,2$, containing $N_k$ neurons respectively, and consider the compound system 
$S=(S_1,S_2)$. In quantum theory and GPT this system should be described by the tensor product construction based  on the 
linear space $M\equiv M_{N_1 N_2} =M_{N_1} \otimes M_{N_2}$. 
How can one couple this tensor product construction to our neuronal networks model? Straightforwardly one treats $S$ as the network with $N_A+N_B$ communicating neurons and represents $S$ by  
matrix of the size  $(N_1+N_1)^2$, and not of the size  $(N_1 N_2)^2$ as GPT requieres.     The essence of our further consideration is in understanding that the state of a compound system should be interpreted differently from the state of a single system. This is one of the consequences 
the studies on so called {\it prequantum classical statistical field theory} \cite{Beyond}, an attempt to model quantum phenomena with classical random fields. In this article we proceed with this ideology which matches well with our neuronal model.  

We turn to the concrete representation of the weights $w_{ij}$ via averages of the classical random variables $\xi_{ij}$, 
see (\ref{xi1}), (\ref{xi1a}). Denote these variables for networks $S_1,S_2$ as $\xi^{(1)}_{ij}, \xi^{(2)}_{ij}$. Consider the covariance matrix for these variables 
\begin{equation}
\label{RV1a}
w_{ijkm}= E[\xi^{(1)}_{ij} \xi^{(2)}_{km}]. 
\end{equation}
The matrix $W_S=(w_{ijkm})$ belongs to the matrix space $M\equiv M_{N_1 N_2}= M_{N_1}\otimes M_{N_2}$. 
In this case, $w_{ijkm}$ represents the joint probability of the event 
that both  $O^{(1)}_{i \to j}$ and $O^{(2)}_{k \to m}$ occur simultaneously, i.e.,
\begin{equation}
\label{RV1ab}
w_{ijkm}=P(\xi^{(1)}_{ij}=1,\xi^{(2)}_{km}=1).
\end{equation}

We remark that the basis of linear space $M$ can be selected as $E_{ijkm}= E_{ij}\otimes E_{km}$. Each $W \in M$, can be represented as the linear combination 
$W= \sum_{ijkm} w_{ijkm} E_{ijkm}$. The norm on $M$ is defined as $||W||= \sum_{ijkm} |w_{ijkm}|$. A matrix $W$ is positive if 
all its elements are non-negative, i.e., the positive cone $C_+=\{W: w_{ijkm} \geq 0\}$. So, matrices generated by the compound neuronal network via (\ref{RV1a}) are positive. The state space of the compound system $S=(S_1,S_2)$ is defined as in (\ref{norm1}),
$\Omega=\{\omega: \omega \geq 0, ||\omega||=1 \}$. The matrix $W_S$  of  the network $S$, see (\ref{RV1a}), is transformed to the corresponding state via normalization, $W_S \to \omega= W_S/||W_S||$. 

Thus, the GPT description of the compound network via the covariance matrix 
differs from the GPT description  of the network $S_1\cup S_2$ composed of all neurons belonging to $S_1$ and $S_2$ and connected by the directed edges .   For   $S_1\cup S_2$ description,  we need to establish all connections between the neurons $n_i^{(1)} \in S_1$ and $n_j^{(2)}\in S_2$ and assign to the corresponding directed edges the weights. The $S_1\otimes S_2$ description doesn't contain these connections between neurons of $S_1$ and $S_2$. Only the cross-correlations between signals inside these networks are taken into account. In this sense  
the $S_1\otimes S_2$ description is incomplete, since it doesn't  recount connections between the neurons of the networks $S_1$ and $S_2$. On the other hand, it recounts correlations, or more specifically the concurrences, between signals within the network $S_1$ and  the network $S_2$.

\begin{figure}
\begin{center}0,1; 
 \includegraphics[width=1\textwidth]{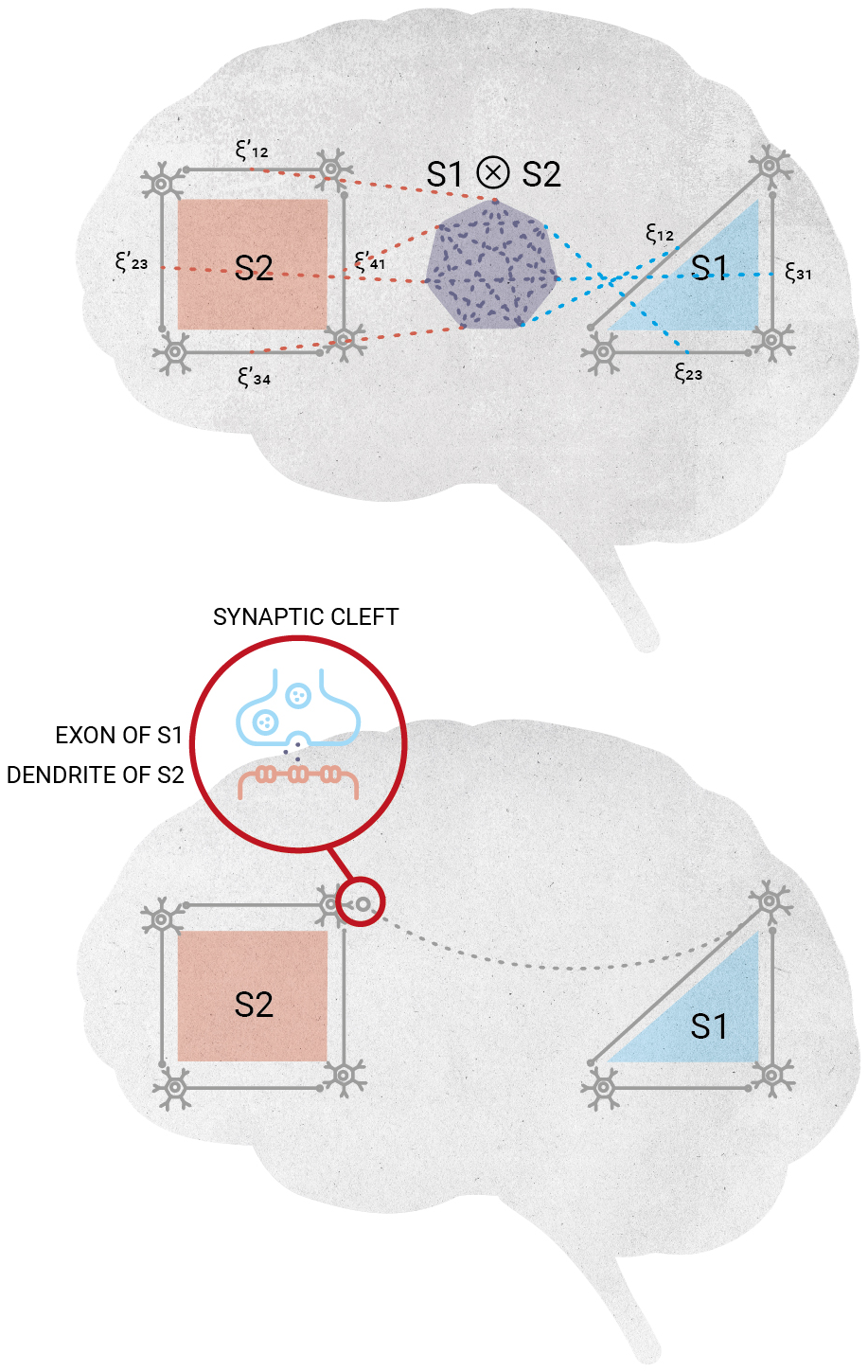}\hspace{8mm} 
 \end{center}
\label{Tensor}
\caption{Visualization of correlations between signaling within two networks $S_1$ and $S_2$ (top panel) and 
the direct physical connection between them (bottom panel), see Remark for further clarification and discussion. } 
\end{figure}

The graph of the $S_1\otimes S_2$ description can be constructed as follows. Denote the edges of the directed graphs $G_{S_p}, p=1,2$, by the symbols
$e^{(p)}_{ij}$. These are the vertexes of the  graph under construction that is symbolically denoted as $G_{S_1\otimes S_2}$.
This graph is undirected and weighted,  and any vertex $e^{(1)}_{ij}$ is connected with any vertex $e^{(2)}_{km}$ 
by the edge $e^{(12)}_{ijkm}$ with the weight $w_{ijkm}$. For example, for $N_1=2, N_2=3$, the graph $G_{S_1\otimes S_2}$ has 
$32$ (undirected) edges and the graph $G_{S_1\cup S_2}$  has 25 (directed) edges. The graph  $G_{S_1\otimes S_2}$ is not the standard tensor product $G_{S_1} \otimes G_{S_2}$ of the graphs $G_{S_1}$ and $G_{S_2}$. The latter graph will be used for the description of compound systems in models 2, 3. The graph  $G_{S_1\otimes S_2}$ can be called the {\it edge tensor product} of the directed graphs $G_{S_1}$ and $G_{S_1}$.  Since in such a graph the structure of the original neuronal 
graphs $S_1$ and $S_2$ would be shadowed, we illustrate the tensor product description of the compound system $S=(S_1,S_2)$ by the neuronal graphs for $S_1$ and $S_2$ and correlation coupling between them, so  see Fig. \ref{Tensor} for such vizualization

\newpage

\newpage

{\bf Remark.} \begin{footnotesize}
 Each of two networks $S_1$ and $S_2$ is characterized by internal signaling, described by the random signals 
$\xi_{ij}$ in $S_1$ and $\xi_{ij}^\prime$ in $S_2$ (for  simplicity of visualization some possible connections 
between neurons are omitted). This is the direct signaling within the networks. Besides this direct signaling, there 
exist temporal concurrences between the signals in $S_1$ and $S_2$ (propagating along the edges in each of them). Mathematically 
they can be represented as correlations between the random signals $\xi_{ij}$ and $\xi_{ij}^\prime,$ see (\ref{RV1a}).                   
 These correlations  are basic for the quantum-like representation via the tensor product. Of course, there can exist direct physical connections between the neurons from $S_1$ and $S_2$ (see bottom panel of Fig. \ref{Tensor}). In the tensor product representation  such direct signaling between the neurons from these two networks are ignored. The latter can be justified only if the number of such direct $S_1\leftrightarrow S_2$ connections is very small compare to the numbers of internal connections within
$S_1$ and $S_2.$ Otherwise the correlation matrix would not reflect the functioning of these two networks. Hence, ``networks entanglement'' physically means very dense connectivity (intensive signaling) within each of them in combination with poor inter-network connectivity (weak inter-network signaling), cf. with \cite{Scholes1}. The crucial question is to understand how the compound network can explore the correlations between signals within them - in the absence of physical connections. We can only speculate that the electromagnetic field within the brain should be involved in establishing of such inter-network and generally long-distance correlations.
\end{footnotesize}

Suppose now that signals in $S_1$ and signals in $S_2$ are independent, i.e., the corresponding random variables are independent.
Then $w_{ijkm}= E[\xi^{(1)}_{ij} \xi^{(2)}_{km}] = E[\xi^{(1)}_{ij}] E[ \xi^{(2)}_{km}] =w^{(1)}_{ij} w^{(2)}_{km}$ that is 
$W_S= W_{S_1} \otimes W_{S_2}$. We also note that $||W||= ||W^{(1)}|| \; ||W^{(2)}||$ and hence the state of $S$ is factorized,
$\omega= \omega^{(1)} \otimes \omega^{(2)}$. Here  $\omega^{(1)}$ and $\omega^{(2)}$ are the states of the networks 
$S_1$ and $S_2$, $\omega^{(1)}= \omega_{S_1}$ and $\omega^{(2)}=\omega_{S_2}$, as they were defined in section \ref{model}. 

For a state $\omega \in \Omega_S$, the states of its subsystems are defined as follows,
$$
\omega^{(1)}= (\omega^{(1)}_{ij}=   \sum_{km} w_{ijkm}), \; \omega^{(2)}= (\omega^{(1)}_{km}=   \sum_{ij} w_{ijkm}),
$$
and $\omega^{(p)} \in \Omega_{S_p}, p=1,2$. We  remark that in terms of random signaling variables, for example,    
$\omega^{(1)}_{ij}= E[\xi^{(1)}_{ij} (\sum_{km} \xi^{(2)}_{km})]$, and the random variable $\Xi^{(2)}=\sum_{km} \xi^{(2)}_{km}$ 
can be interpreted as the intensity of signaling within the network $S_2$.  Hence, the network $S_1$ can be considered as open quantum-like system interacting with environment - the network $S_2$, and vice verse. In this model $S_1$ ``feels'' not signaling between the individual neurons of $S_2$, but the integral intensity of signaling.        

One can  proceed further by employing the general formalism of measurement instruments in tensor products. 

\section{Concluding remarks}

As noted, quantum-like modeling is actively developing as a phenomenological approach. However, it strongly requires integration with brain functioning, specifically with the activity of networks of communicating neurons. In this article, we established the basics of such integration within the GPT model, which represents networks using ordered weighted graphs. These graphs are described by weight matrices that quantify the intensities of signaling between neurons. The graph-matrix GPT model (Model 1) has a classical state space — a polygon — and its quantum-like properties are generated by the calculus of measurement instruments.

We presented several quantum-like effects — such as the order, interference (disjunction), and non-repeatability effects — within the graph-matrix GPT model. The discussion of compound systems and correlated neuronal networks was brief, and we plan to explore the quantum-like properties related to the entanglement of neuronal network states in a future publication.

\section*{Acknowledgments} 
A.K. was partially supported by  the EU-grant CA21169 (DYNALIFE), by  JST, CREST Grant Number JPMJCR23P4, Japan, and visiting professor fellowship to Tokyo University of Science (April 2024).
M.O. was partially supported by JSPS KAKENHI Grant Numbers JP24H01566, JP22K03424, JST CREST Grant Number JPMJCR23P4,  
RIKEN TRIP Initiative (RIKEN Quantum), and  the Quantinuum--Chubu University Collaboration in 2023--2024. A.K. would like to thank 
prof. Bussmeyer for supporting discussion on the

\end{document}